# Surface morphology, structure and transport property of $Na_xCoO_2$ thin films grown by pulsed laser deposition

X. P. Zhang, Y. S. Xiao, H. Zhou, B. T. Xie, C. X. Yang, and Y. G. Zhao

*Department of Physics, Tsinghua University, Beijing 100084, P. R. China*

**Abstract.** In this paper, we report the growth of $Na_xCoO_2$ thin films by pulsed-laser deposition (PLD). It is shown that the concentration of sodium is very sensitive to the substrate temperature and the target-substrate distance due to the evaporation of sodium during the deposition. α '-phase $Na_{0.75}CoO_2$ and γ- phase $Na_{0.71}CoO_2$ thin films can be obtained with different conditions. Correspondingly, the surface morphology of the films changes from flake-like to particle-like. The temperature dependence of resistivity for the films prepared with the optimal condition shows metallic behavior, consistent with the data of $Na_xCoO_2$ single crystals. This work demonstrates that PLD is a promising technique to get high quality $Na_xCoO_2$ thin films.

**Introduction**
Recently the discovery of superconductivity in water-intercalated $Na_xCoO_2$ compound has attracted much attention [1]. The parent compound $Na_xCoO_2$ itself is also attractive due to its unique structure and physical properties, e.g., large room temperature Seebeck coefficient 100 μv/K, which is nearly ten times higher than that of typical metals [2]. It also has very low resistivity and its carrier concentration is much higher than that in some known thermoelectric materials like $Bi_2Te_3$ and PbTe [3]. It is believed that the large thermoelectric power in $Na_xCoO_2$ can not be accounted by conventional one electron picture and spin entropy or spin fluctuation may play an important role [2, 4]. Similar to high $T_c$ superconductors and manganites, $Na_xCoO_2$ also belongs to the strongly correlated systems [5, 6]. The fabrication of high quality $Na_xCoO_2$ thin films is important for both basic research and device applications. Up to now, to our knowledge, no reports have been seen in the literature. In this paper, we report the growth of $Na_xCoO_2$ thin films by pulsed-laser deposition and their properties.

**Experimental**
The targets with nominal compositions of $Na_{0.75}CoO_2$ was prepared by solid state reaction using $Co_3O_4$ and $Na_2CO_3$. It was sintered at $860^oC$ for 12 hours in the furnace in air. By X-ray diffraction, the pellet sample was confirmed to be single phased hexagonal γ-$Na_xCoO_2$ without any traces of impurity phases. The energy spectrum analyzer shows that the sample is roughly stoichiometric – that is Na/Co = 0.68. This result indicated that the Na content was less than the original stoichiometry due to the evaporation of sodium during sintering.

The sodium cobalt oxide thin films were prepared by Pulse Laser Deposition (PLD). The film was deposited on $LaAlO_3$ (LAO) (001) substrates. The laser energy density at the target was about

3 J/cm$^2$ at a laser flux of 800mJ/pulse. The substrate temperature was tuned in order to optimise the conditions. The deposition pressure in the chamber was 38 to 40 Pa with pure oxygen environment and the substrate was placed in front of the target at a distance of 3.8 to 4.5 cm. The films were deposited at a repetition rate of 5 Hz for 5 minutes, yielding an approximate layer thickness of 200 nm. The resistivity measurements were carried out using the DC four probe methods. The structures were analyzed using an x-ray diffractometer (XRD). The surface morphologies were shown by scanning electron microscope (SEM) and the Na/Co ratio was determined by EDAX. The Seeback coefficient was measured in air from room temperature to 250°C.

**Results and discussion**
Since sodium is easy to evaporate during deposition, we carefully tuned the substrate temperature and the substrate-target distance. It was found that high substrate temperature (>680 °C) leads to the dramatic loss of sodium and the films mainly contain some impurity phases and the films are insulating, while low substrate temperature hinders the formation of Na$_x$CoO$_2$ phase due to the low kinetic energy of the deposited atoms. Large substrate-target distance is also unfavorable for the formation of Na$_x$CoO$_2$ phase. Figure 1 shows the typical x-ray diffraction patterns for samples A, B and C. Sample A was grown at 650 °C with about 4.2 cm substrate-target distance. Sample B was also grown at 650 °C but with 4.5 cm substrate-target distance. Sample C was grown at 670 °C with about 4.2 cm substrate-target distance. Na$_x$CoO$_2$ compounds were hexagonal or monoclinic structured with different Na contents (0.5< x < 1). There is a series of early works studying Na$_x$CoO$_2$ with various x of sodium [7-9]. Four relevant phases α (0.9< x < 1.0), α'(x = 0.75), β(0.55 < x < 0.60), and γ(0.50 < x < 0.75) can exist [7]. The β phase structure is different from the γ phase in which the oxygen atoms of prism shared different CoO$_6$ octahedra [10]. By comparison with the results in reference [7], it can be deduced that sample A is nearly single phased γ-Na$_{0.71}$CoO$_2$, while sample B and C contains quite some impurity phases as indicated in Fig.1.

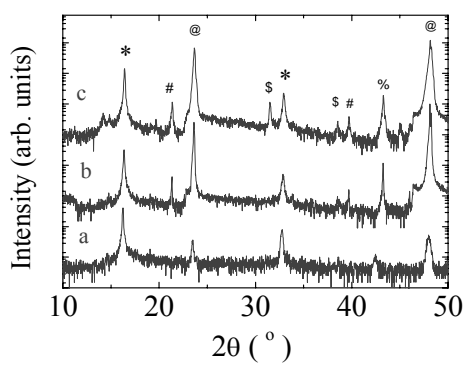
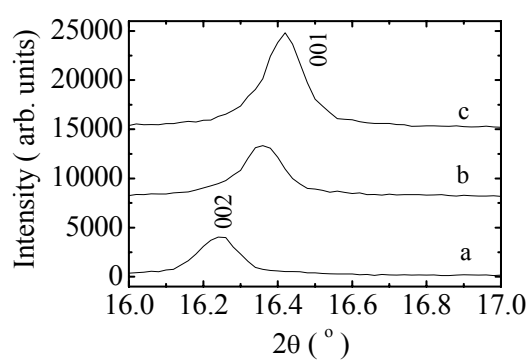

Fig.1 X-ray diffraction patterns for Na$_x$CoO$_2$ thin films prepared with different conditions. Peaks denoted by * belong to Na$_x$CoO$_2$, corresponding to (002), (004) for sample A and B and (001), (002) for sample C. @ stands for LAO substrate peaks. #, $ and % are Na$_3$CoO$_2$, Co$_3$O$_4$ and CoO peaks, respectively.

Fig.2 Peak position variation in x-ray diffraction patterns for Na$_x$CoO$_2$ thin films prepared with different conditions.

It is also noted that the peak position shifts to high angles from sample A to sample C as shown in Fig. 2 and changes from γ-Na$_{0.71}$CoO$_2$ phase to α'-Na$_{0.75}$CoO$_2$ phase. For γ-Na$_{0.71}$CoO$_2$, it has hexagonal structure and the x-ray diffraction peak at about 16.25 ° is (002) peak. For α'-Na$_{0.75}$CoO$_2$, it has the monoclinic structure and the x-ray diffraction peak at about 16.45 ° is (001) peak. From these angles, we can obtain the lattice parameter $c$ for samples A, B, and C as shown in Table 1.

Table 1 Lattice parameter $c$, the ratio of Na/Co and resistivity at room temperature for samples A, B and C.

| Samples | $c$ (Å) | Na / Co | ρ(T=300K) (mΩ· cm) |
|---|---|---|---|
| A | 10.915 | 0.6 | 1.69 |
| B | 10.836 | 0.3 | 1240 |
| C | 5.794 | 0.5 | 16.62 |

Figure 3 is the surface morphology of samples A, B and C. Sample A shows rather compact flake-like crystals, which should be Na$_x$CoO$_2$. For sample B, there are a lot of small grains in the regions between the flake-like crystals. These small grains are presumably related to the impurity and block the connection between the flake-like Na$_x$CoO$_2$ crystals, leading to the dramatic increase of resistivity at room temperature as shown in Table 1. For sample C, the surface morphology shows three dimensional crystals and random alignment, which is totally different from the flake-like crystals of sample A and B. This morphology difference may be related to the difference of crystal structures for γ-Na$_{0.71}$CoO$_2$ phase to α'-Na$_{0.75}$CoO$_2$, The two dimensional nature may be more strong in γ-Na$_{0.71}$CoO$_2$ phase than that in α'-Na$_{0.75}$CoO$_2$. The average Na/Co ratios for samples A, B and C are also shown in Table 1. Compared with the room temperature resistivity in Table 1, it is noted that large Na/Co ratio corresponds to low resistivity. This correlation can be understood by considering the phase purity and surface morphology of the Na$_x$CoO$_2$ thin films.

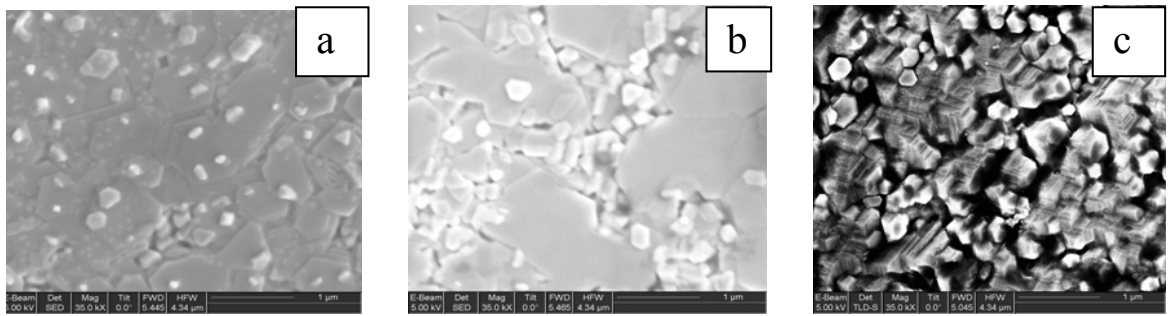

Fig.3 Surface morphology for Na$_x$CoO$_2$ thin films prepared with different conditions.

Shown in Fig.4 is the electrical resistivity for samples A, C with respect to temperature. The resistivity has been normalized by room temperature resisitvity. The corresponding room temperature resistivity can be found in Table 1. The behavior of sample A is consistent with that of Na$_x$CoO$_2$ single crystals and shows metallic transport property. It is interesting to note that the resistivity of sample C shows a linear temperature dependence from room temperature to 5 K. This behavior has not been reported in the literature. This unusual behavior deserves further study.

Figure 5 is the temperature dependence of Seebeck coefficient for sample C above room temperature. The Seebeck coefficient shows linear increase with temperature and is consistent with that of the bulk $Na_xCoO_2$ samples.

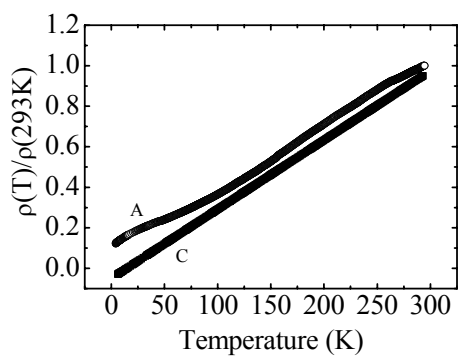
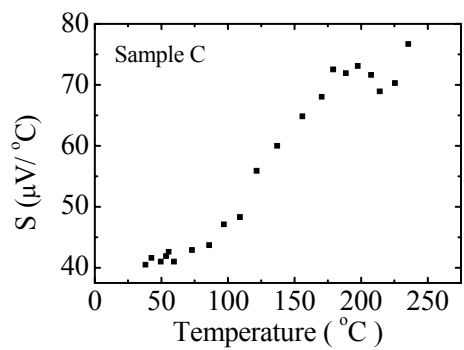

Fig.4  Electrical resistivity ρ for the $Na_xCoO_2$ samples with respect to temperature.

Fig.5 Variation of Seeback coefficient with temperature for sample C.

In summary, we have prepared $Na_xCoO_2$ thin films by PLD. By tuning deposition condition, γ-$Na_{0.71}CoO_2$ phase and α'-$Na_{0.75}CoO_2$ phase were obtained. The surface morphology of γ-$Na_{0.71}CoO_2$ phase corresponds to compact flake-like crystals while the α'-$Na_{0.75}CoO_2$ phase shows gathering of small grains. The electrical transport property of the $Na_xCoO_2$ thin film, prepared with optimal condition, is consistent with that of $Na_xCoO_2$ single crystals.

**Acknowledgements**
This work was supported by the Excellent Young Teacher Program of MOE, P.R.C, Specialized Research Fund for the Doctoral Program of Higher Education (No. 2003 0003088), NSFC (project No. 50272031) and National 973 project (No. 2002CB613505).